\documentclass{article}

\newtheorem{remark}{Remark}

\usepackage{graphicx}
\usepackage{bbm}
\usepackage{caption}
\usepackage{subcaption}
\usepackage{amssymb}
\usepackage{amsmath}
\usepackage{amsfonts}
\usepackage{xcolor}
\usepackage{lineno,hyperref}
\usepackage{epstopdf}

\begin{document}

\title{Bifurcation analysis and steady state patterns in reaction-diffusion systems augmented with self- and cross-diffusion.}

\author{Benjamin Aymard}

\maketitle

\begin{abstract}

In this article, a study of long-term behavior of reaction-diffusion systems augmented with self- and cross-diffusion is done,
using an augmented Gray-Scott system as a generic example.
The methodology remains general, and is therefore applicable to other systems.
Simulations of the temporal model (nonlinear parabolic system) reveal the presence of steady states, often associated with energy dissipation.
A Newton method based on a mixed finite element method is provided, 
in order to directly evaluate the steady states (nonlinear elliptic system) of the temporal system, 
and validated against its solutions.
Linear stability analysis (LSA) using Fourier analysis is carried out around homogeneous equilibrium, 
and using spectral analysis around non homogeneous ones. For the latter, the spectral problem is solved numerically.
A multi-parameter birfurcation is reported.
Original steady state patterns are unveiled, not observable with linear diffusion only.
Two key observations are made: a dependency of the pattern with the initial condition of the system, 
and a dependency on the geometry of the domain.
\end{abstract}

\section{Introduction}

Reaction-diffusion systems are ubiquitous, and widely appear in nature,  
from physics (fluid instability, reacting flow), to chemistry (autocatalytic reaction, flame propagation) to biology (population dynamics, excitable media).
These systems can generate, under certain conditions, different kinds of instabilities 
(\cite{Manneville2004}, \cite{Perthame2015}), 
sometimes leading to the formation of structures known as Turing patterns (\cite{Turing1952}),
first observed experimentally in the so-called Chloride-Iodide-Malonic Acid (CIMA) reaction (\cite{Kepper1990}).
Linear stability analysis (\cite{Kuznetsov1998}) is an efficient tool for predicting the critical value of parameters which can trigger instabilities in the system.
Alternatively, a Lyapunov stability analysis may be done, such as \cite{You2007} and \cite{Lombardo2008}.
Such an energetic approach is advantageous, when applicable, as it does not necessarily rely on a local linearization of the dynamics.
Although most systems consider two species, models with three or more species are also relevant, for instance in population dynamics (food-chain modelling \cite{Abid2015}) or chemical reaction  
(Belousov–Zhabotinsky reaction dispersed in water-in-oil aerosol experiments \cite{Epstein2001}).

Reaction-diffusion systems augmented with self- and/or cross-diffusion terms have been an active research topic since the seminal work of \cite{SKT1979} in population dynamics, where a generalization of a predator-prey model was introduced, called the SKT system. 
Cross-diffusion also naturally appears in chemistry, where it is experimentally observed in various systems such as micro-emulsions, mycelles, electrolytes, and often modelled with the Steffan-Maxwell model of multi-component diffusive transport (see for instance \cite{Cances2020} for a recent numerical analysis of such systems).
Pattern formation, often related to Turing instabilities, has been reported in various systems containing self- and/or cross-diffusion \cite{Epstein2008}, \cite{Gambino2012}, \cite{Yin2013}, \cite{Li2019}, \cite{Moussa2019}, \cite{Breden2021}, \cite{Zhang2021}.
Recently, Hopf instabilities were also reported in the framework of the SKT system (\cite{Soresina2022}). 

A classical example of a reaction-diffusion system is the Gray-Scott (GS) system \cite{GS1983}, 
modeling an autocatalytic process between two chemical species in a stirred tank.
This system is a generalization of the differential system of Selkov, modeling the glycolysis process \cite{Selkov1968},
and is known to be able to generate a wide variety of patterns, 
observed both in simulations (\cite{Pearson1993}), and in actual experiments on Ferrocyanide-Iodate-Sulfite reaction (\cite{Lee1994}).

As a generic example, we study a modified version of the Gray-Scott system,
augmented with self- and cross-diffusion terms, introduced in \cite{Aymard2022}.
Original transient patterns have been reported in this reference, not observed with linear diffusion only.
One notable finding was that, unlike classical reaction-diffusion systems,
different patterns may be observed by varying only the nonlinear diffusion coefficients, for the same reaction term.
This article also introduced an energy law for this type of system,
and preliminary estimations of energy evolution were showing a tendency for energy dissipation, towards a possible steady state.

In this article, we continue this work, studying the long-term behavior of the system,
by evaluating the equilibrium solutions, and their local stabilities.
Evaluation of the steady states is a challenging problem, as one needs to solve a system of nonlinear partial differential equations,
possibly within a complex geometry, that may dramatically modify the observed patterns. 
Stability analysis is another challenging problem: 
Fourier analysis is preferable around homogeneous steady states, however numerical methods solving a spectral problem is more useful around non homogenous ones. 

This article is organized as follows.
In the first section, steady states are studied. 
Starting with numerical simulations of the energy evolution, revealing the presence of stable equilibria,
a direct method to compute the steady states is introduced and validated against the long-term solutions of the temporal model.
The second section is devoted to the stability analysis of the steady states.
First, by using Fourier analysis, leading to a marginal stability surface and corresponding critical wavelengths.
Then, with a numerical approach, consisting of evaluating the eigenvalues of the linearized PDE around a non homogeneous steady state.
The last section presents a numerical exploration of the patterns and exploring various geometries.

\section{Steady states}

In this section we study the long-term behavior of reaction-diffusion systems augmented with self- and cross-diffusion, that reads:
\begin{align}
\frac{\partial \phi_1}{\partial t} &= \Delta \mu_1 + R_1,\notag\\
\frac{\partial \phi_2}{\partial t} &= \Delta \mu_2 + R_2,\label{dGS}
\end{align}
with $R_1, R_2$ the reaction terms and $\mu_1, \mu_2$ chemical potentials, defined by:
\begin{align}
\mu_1 &= (d_1 + d_{11}\phi_1^{2} + d_{12}\phi_2^{2})\phi_1,\notag\\
\mu_2 &= (d_2 + d_{22}\phi_2^{2} + d_{21}\phi_1^{2})\phi_2.\label{mGS}
\end{align}
The problem is closed by adding initial conditions:
\[
(\phi_1(t=0),\phi_2(t=0))= (\phi_1^0,\phi_2^0),
\]
and null-flux boundary conditions: 
\[
\frac{\partial \phi_1}{\partial n} = \frac{\partial \phi_2}{\partial n} = 0.
\]
The variables $d_1, d_{11}, d_2, d_{22}, d_{12}, d_{21}$ are real non-negative parameters.
In the context of classical reaction-diffusion systems (for which only linear diffusion is considered, $d_{11} = d_{12} = d_{22} = 0$), 
$\phi_1$ is called an inhibitor, and $\phi_2$ an activator, and it has been showed that, under condition, this model presents a Turing instability, 
when a slow diffusion of the activator is coupled to a rapid diffusion of the inhibitior.
In the rest of the article, we will only consider the symmetric case, for which $d_{21} = d_{12}$.
As a generic example, we will consider along the article, the reaction terms corresponding to the Gray-Scott model, defined by:
\begin{align}
R_1 &= - \phi_1\phi_2^2 + F(1-\phi_1),\notag\\
R_2 &= \phi_1\phi_2^2 - (F+k)\phi_2,\notag
\end{align}
with $F, k$ real non-negative parameters.

\begin{remark}
In this article, only a chemical potential of the form \eqref{mGS} is considered, but the methodology that is developped is applicable to more general cases,
for instance polynomials, as long as the coefficients remain all non-negative.
\end{remark}

\subsection{Energy dissipation}

\begin{figure}[!htbp]\centering
\begin{subfigure}{0.4\textwidth}
\includegraphics[width=\linewidth]{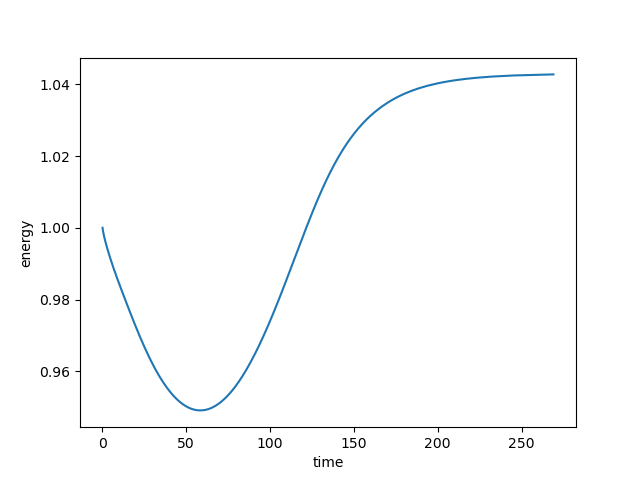}
\caption{Homogeneous case
\\$d_1 = 1e^{-5}, d_2 = 1e^{-5},
\\d_{11} = 0, d_{22} = 0, d_{12}=0$.}
\end{subfigure}
\begin{subfigure}{0.4\textwidth}
\includegraphics[width=\linewidth]{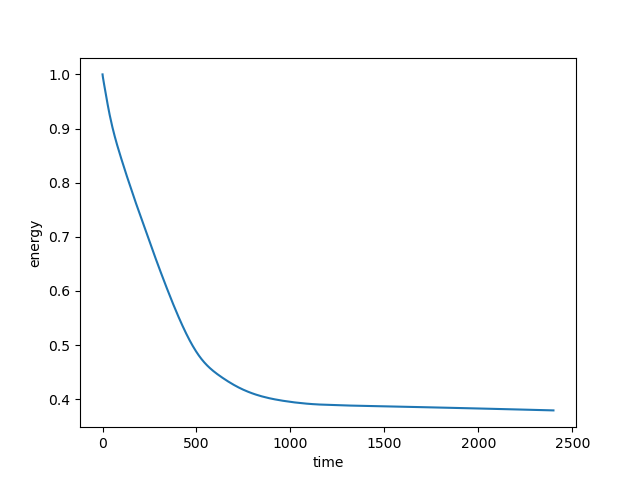}
\caption{Linear diffusion instability
\\$d_1 = 2e^{-5}, d_2 = 1e^{-5},
\\d_{11} = 0, d_{22} = 0, d_{12} = 0$.}
\end{subfigure}
\begin{subfigure}{0.4\textwidth}
\includegraphics[width=\linewidth]{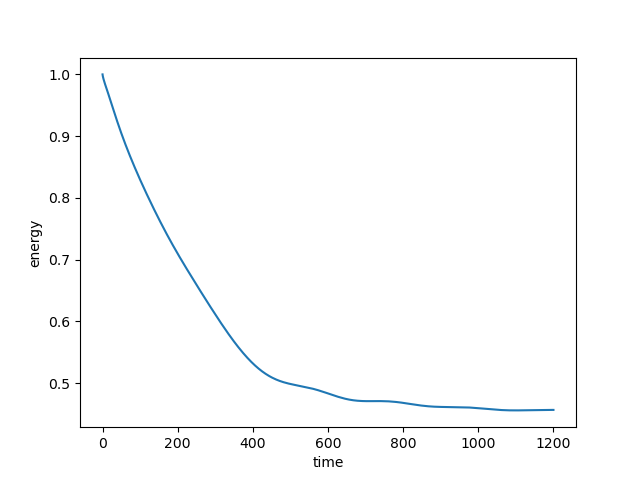}
\caption{Self-diffusion instability
\\$d_1 = 2e^{-5}, d_2 = 1e^{-5},
\\d_{11} = 5e^{-6}, d_{22} = 0, d_{12} = 0$.}
\end{subfigure}
\begin{subfigure}{0.4\textwidth}
\includegraphics[width=\linewidth]{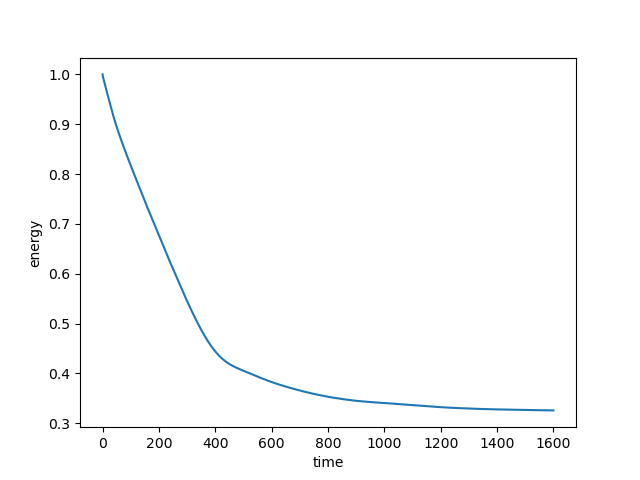}
\caption{Linear and self- and cross-diffusion instability
\\$d_1 = 2e^{-5}, d_2 = 1e^{-5},
\\d_{11} = 2e^{-6}, d_{22} = 0, d_{12}=1e^{-6}$.}
\end{subfigure}
\caption{Evolution of energy (\ref{energy}) with respect to time for four key cases
(the ordinate axis is scaled with initial energy, in order to always start at $E=1$),
obtained by solving the temporal model (\ref{dGS}, \ref{mGS}), starting from initial condition \eqref{CI},
and null flux boundary conditions.
We observe in each case a convergence towards an equilibrium state.}
\label{Ecurve}
\end{figure}

Systems of the form (\ref{dGS}, \ref{mGS}) respect the following energy law \cite{Aymard2022}:
\begin{equation}
\frac{d}{dt}E(\phi_1,\phi_2) = - \| \nabla \mu_1 \|^2 - \| \nabla \mu_2 \|^2 + (R_1,\mu_1) + (R_2,\mu_2), 
\label{Elaw}
\end{equation}
with $E$ defined by:
\begin{equation}
E(\phi_1,\phi_2) = 
\int_{\Omega} 
\left(d_1 \frac{\phi_1^2}{2} + d_{11}\frac{\phi_1^4}{4} 
+ d_2 \frac{\phi_2^2}{2} + d_{22}\frac{\phi_2^4}{4} 
+ d_{12}\frac{\phi_1^2 \phi_2^2}{2}
\right) d\omega .
\label{energy}
\end{equation}
In order to study its long-term behavior, model (\ref{dGS}, \ref{mGS}) was solved numerically,
monitoring energy (\ref{energy}), looking for potential convergence towards a steady state.
The numerical method provided in the cited reference was used, and implemented on FreeFem++ software \cite{Hecht2012}.
Let us recall that the scheme is second order accurate in time, and second order accuracy in space is reached by using P2 finite element method.
In this case, adaptive mesh refinement strategy was used, with mesh size varying from $h_{\min} = 0.001$ to $h_{\max} = 0.07$,
according to both the variation of $\phi_1$ and $\phi_2$, and the time step was set to $\Delta t = 0.1$.
Model parameters were set to: $F = 0.037$, $k = 0.06$, within a square domain $\Omega = [-0.2,0.2]^2$, starting from initial condition:
\begin{align}
\phi_1^0 &= (C + 0.05U)\mathbf{1}_S + \mathbf{1}_{\Omega \setminus S}, \label{CI}\\
\phi_2^0 &= (0.25 + 0.05V)\mathbf{1}_S, \notag
\end{align}
with $C = 0.3$, $S = [-0.05,0.05]^2$, and $U,V$ random numbers uniformly distributed on $[0,1]$.
In Figure \ref{Ecurve}, we plot the energy (\ref{energy}) evolution in four key cases,
representing different combinations of linear and nonlinear diffusion: 
a homogeneous case ($d_1 = 1e^{-5}, d_2 = 1e^{-5}, d_{11} = 0, d_{22} = 0, d_{12}=0$), 
a linearly instable case ($d_1 = 2e^{-5}, d_2 = 1e^{-5}, d_{11} = 0, d_{22} = 0, d_{12} = 0$), 
a self-diffusion instable case ($d_1 = 2e^{-5}, d_2 = 1e^{-5}, d_{11} = 5e^{-6}, d_{22} = 0, d_{12} = 0$), 
and combination of linear, self and cross diffusion case ($d_1 = 2e^{-5}, d_2 = 1e^{-5},d_{11} = 2e^{-6}, d_{22} = 0, d_{12}=1e^{-6}$).
We oberve that, for every case, steady state is reached after a sufficiently long period of time.
Interestingly, energy decreases for every case leading to a pattern formation (dissipative structure),
but increases in the case where a homogeneous state is reached.

\begin{remark}
Let us remark that $E = \int_{\Omega} P(\phi_1,\phi_2)$ with $P$ polynomial with positive coefficients.
Therefore, given that the domain $\Omega$ is finite, $E$ is bounded when densities are bounded. 
Indeed, as an integral of non-negative terms, it is positive:
\[
0 \leq E(\phi_1,\phi_2) \leq |\Omega| P(\|\phi_1\|_{\infty},\|\phi_2\|_{\infty}) < \infty.
\]
This hypothesis is natural, as the associated diffusionless system respect this property,
and as it has been observed numerically in the nonlinear diffusive system. 
However, the proof remains open.
As a consequence, when one considers long time behavior of the system, only three possibilities occur for energy evolution: 
a relaxation towards a steady state, a limit cycle, or a non periodic bounded curve. 
\end{remark}

\begin{remark}
Energy (\ref{energy}) spontaneously decreases in two obvious cases: 
first, when there is no reaction term ($R_i = 0$), and second, if $\forall i, R_i = - \mu_i$.
In general, energy does not necessarily decreases in the Gray Scott case, notably due to the source term $F$,
increasing $R_1$ term in the right hand side of (\ref{Elaw}).
\end{remark}

\begin{remark}
Assuming that $\forall i, \mu_i \in H_0^1(\Omega), R_i \in L^2(\Omega)$, 
a sufficient condition for energy dissipation is to control the reaction term by the dissipative term, using an inequality of the form:
\[
\forall i, \|R_i\| \leq \alpha_R \| \mu_i \|,
\] 
with $\alpha_R \in \mathbb{R}^+$, depending on reacting terms, sufficiently small.
Indeed, using Cauchy-Schwarz and Poincaré inequalities, one gets:
\[
\forall i, 
|(R_i,\mu_i)| \leq \|R_i\| \|\mu_i\| \leq \alpha_R \|\mu_i\|^2 \leq \alpha_R \beta_{\Omega}\| \nabla \mu_i \| ^2,
\]
with $\beta_{\Omega}$,  depending on the geometry of the domain $\Omega$, and therefore:
\[
\frac{d}{dt} E 
= \sum_{i=1}^2 \left(-\| \nabla \mu_i \|^2 + (R_i,\mu_i)\right)  
\leq 
\sum_{i=1}^2 \| \nabla \mu_i \|^2 (\alpha_R \beta_{\Omega} - 1).
\]
Energy dissipation:
\[
\frac{d}{dt} E \leq 0,
\]
is then ensured when $\alpha_R \beta_{\Omega} \leq 1$.
\end{remark}

\subsection{Finite Element Method}

\begin{figure}[!htbp]\centering
\begin{subfigure}{\textwidth}
\includegraphics[width=0.45\linewidth]{Images/linearInstability.png}
\includegraphics[width=0.45\linewidth]{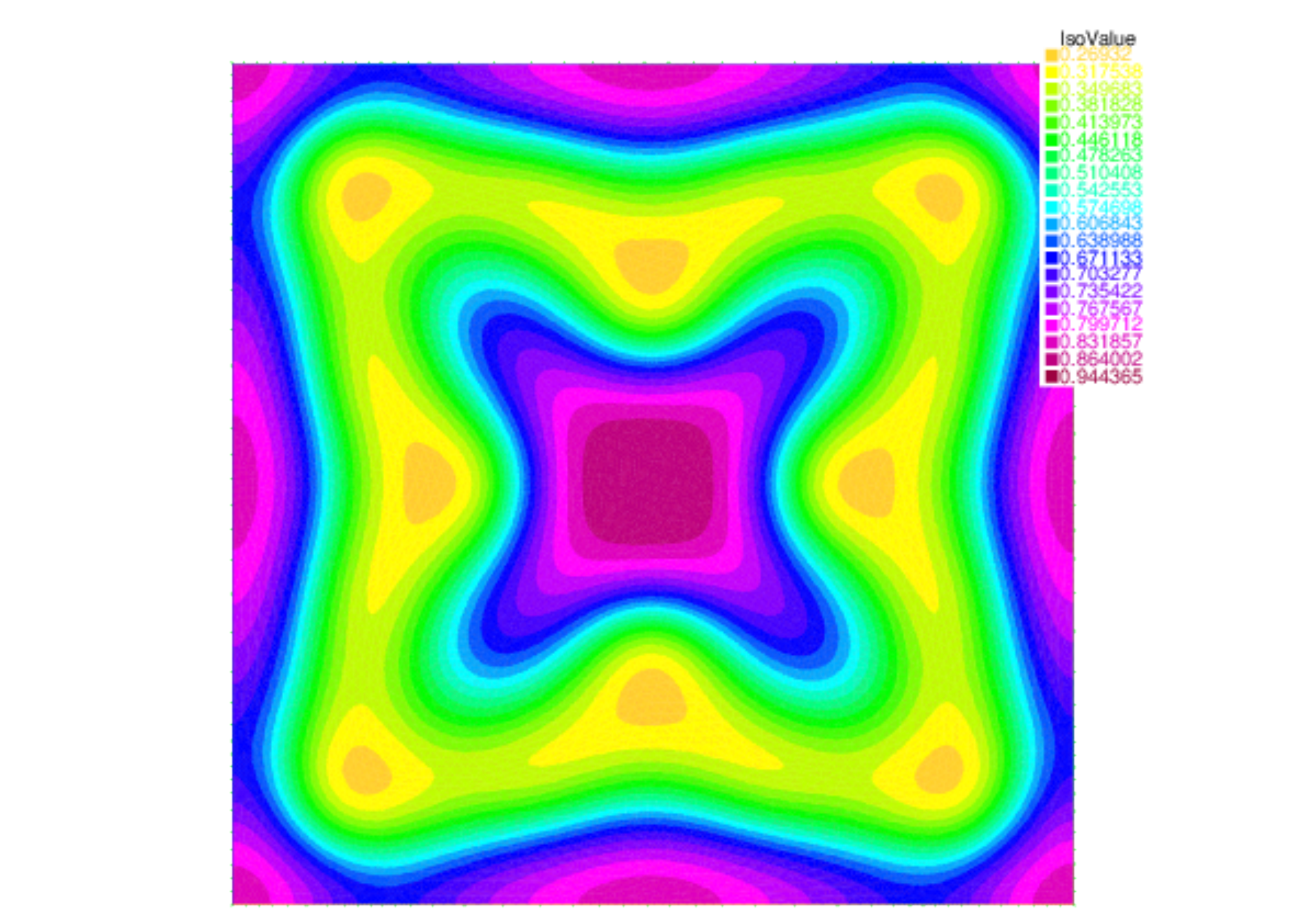}
\caption{Linear diffusion instability:
\\$d_1 = 2e^{-5}, d_2 = 1e^{-5}$,
$d_{11} = 0, d_{22} = 0, d_{12}=0$.}
\end{subfigure}
\begin{subfigure}{\textwidth}
\includegraphics[width=0.45\linewidth]{Images/selfInstability.png}
\includegraphics[width=0.45\linewidth]{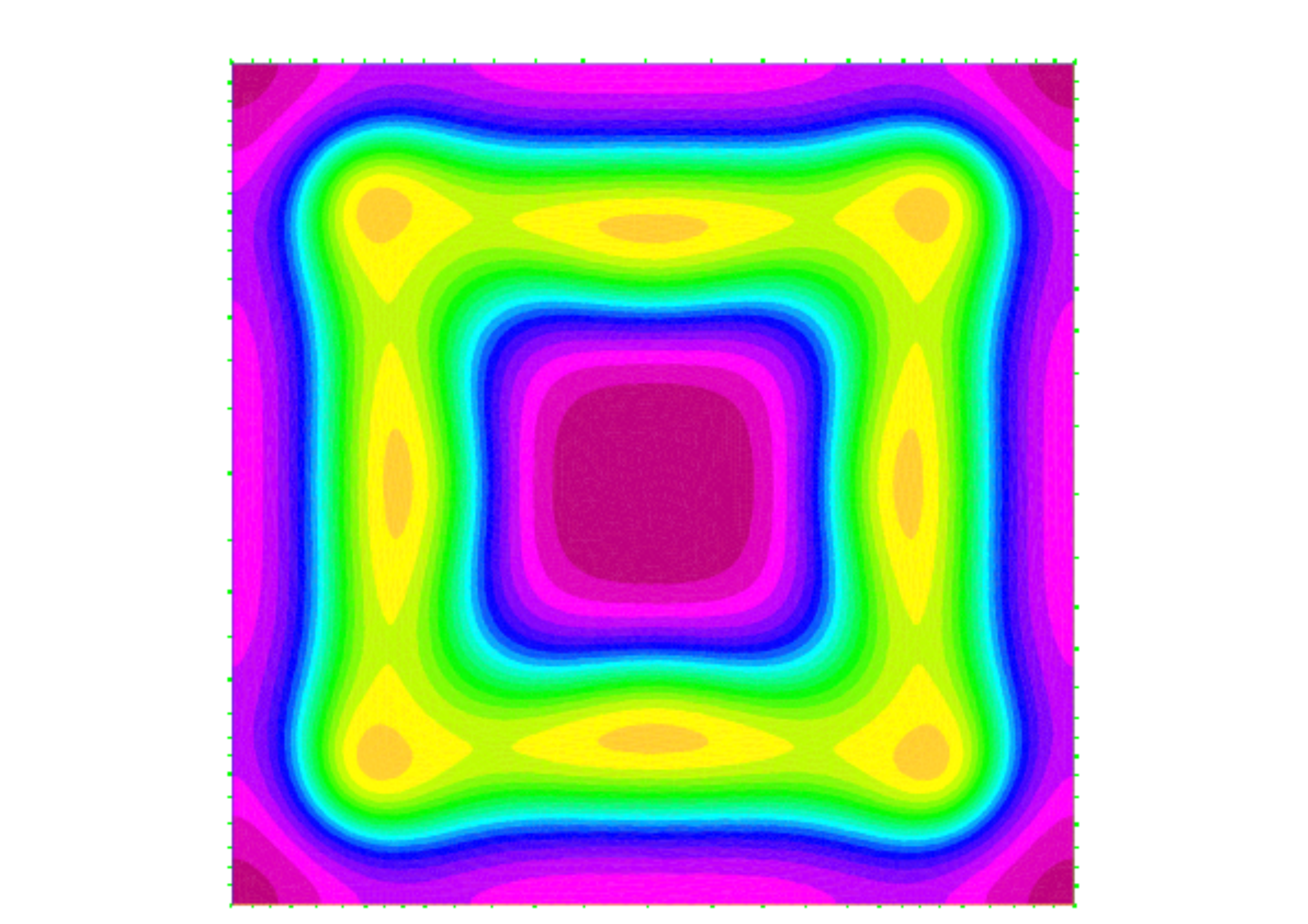}
\caption{Self-diffusion instability:
\\$d_1 = 2e^{-5}, d_2 = 1e^{-5}$,
$d_{11} = 5e^{-6},d_{22} = 0, d_{12} = 0$.}
\end{subfigure}
\begin{subfigure}{\textwidth}
\includegraphics[width=0.45\linewidth]{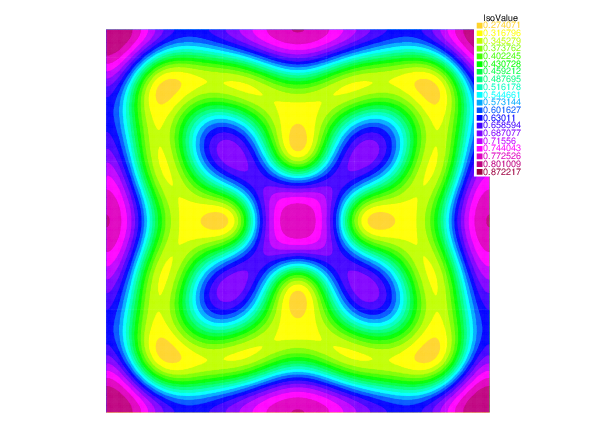}
\includegraphics[width=0.45\linewidth]{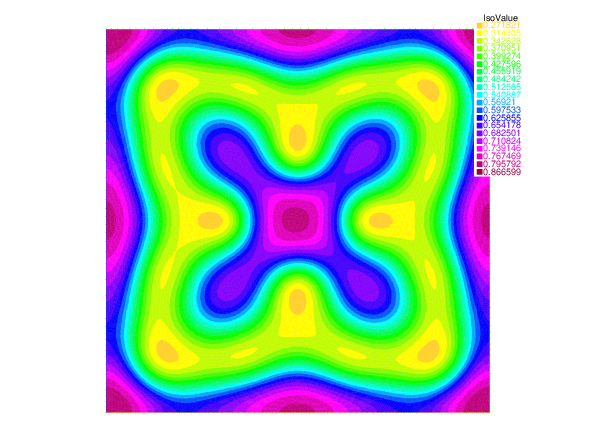}
\caption{Linear and self- and cross-diffusion instability:
\\$d_1 = 2e^{-5}, d_2 = 1e^{-5}$,
$d_{11} = 2e^{-6}, d_{22} = 0, d_{12}=1e^{-6}$.}
\end{subfigure}
\caption{Comparison of steady states of density $\phi_1$ obtained with two different methods:\\
directly solving the steady state problem (nonlinear elliptic system (\ref{NLsteady})) on the left column,
solving the temporal model (nonlinear parabolic system (\ref{dGS}, \ref{mGS})) until equilibirum, on the right column.
A good agreement is observed. The same observation applies on density $\phi_2$, not represented here.}
\label{steadyStates}
\end{figure}

In this subsection, we derive a direct finite element method, based on a Newton algorithm, 
in order to directly compute the steady states of the problem.
Steady states of system (\ref{dGS}, \ref{mGS}), if they exist, verify:
\begin{equation}
R(W) = 0,
\label{NLsteady}
\end{equation}
with:
\[
R(W) = 
\begin{pmatrix}
\Delta \mu_1 + R_1(\phi_1,\phi_2) \\
\Delta \mu_2 + R_2(\phi_1,\phi_2) \\ 
d_1 \phi_1 + d_{11} \phi_1^3 + d_{12}\phi_2^2 \phi_1 -\mu_1  \\
d_2 \phi_2 + d_{22} \phi_2^3 + d_{12}\phi_1^2 \phi_2 -\mu_2  \\
\end{pmatrix},
\]
and $W = (\phi_1,\phi_2,\mu_1,\mu_2)$.
Our strategy to compute numerically the steady states is to apply a Newton method:
\begin{equation}
W^{(n+1)} = W^{(n)} - \delta W,
\label{Newton}
\end{equation}
with $\delta W$ solution of:
\begin{equation}
J_{R(W)} \delta W = R(W).
\label{deltaW}
\end{equation}
Iterations continue until the convergence criteria:
\[
\frac{\| \delta \phi_1\|_{L^\infty}}{\|\phi_1\|_{L^{\infty}}} < \epsilon,
\]
is satisfied. This method requires the knowledge of the Jacobian matrix: 
\[
J_{R(W)}\delta W
=
\begin{pmatrix}
\Delta \delta\mu_1 + \frac{\partial R_1}{\partial \phi_1}\delta\phi_1 + \frac{\partial R_1}{\partial \phi_2}\delta\phi_2\\
\Delta \delta\mu_2 + \frac{\partial R_2}{\partial \phi_1}\delta\phi_1 + \frac{\partial R_2}{\partial \phi_2}\delta\phi_2 \\ 
d_1 \delta \phi_1 + d_{11} 3\phi_1^2\delta \phi_1 + d_{12}(\phi_2^2 \delta \phi_1 + 2 \phi_1 \phi_2 \delta \phi_2) -\delta \mu_1 \\
d_2 \delta \phi_2 + d_{22} 3\phi_2^2\delta \phi_2 + d_{12}(\phi_1^2 \delta \phi_2 + 2 \phi_1 \phi_2 \delta \phi_1) -\delta \mu_2\\
\end{pmatrix}.
\]
Perturbation $\delta W$ is found by solving the following linear system, based on the mixed variational formulation of (\ref{deltaW}):
\begin{multline}\label{FEM}
\forall \psi_1,\psi_2, \nu_1, \nu_2 \in H^1(\Omega),\\
\int_{\Omega} - \left(\nabla \delta\mu_1 , \nabla \psi_1 \right) 
+ \left(\frac{\partial R_1}{\partial \phi_1}\delta\phi_1 + \frac{\partial R_1}{\partial \phi_2}\delta\phi_2\right)\psi_1 
- \left(\nabla \delta\mu_2 , \nabla  \psi_2\right)
+ \left(\frac{\partial R_2}{\partial \phi_1}\delta\phi_1 + \frac{\partial R_2}{\partial \phi_2}\delta\phi_2\right)\psi_2 \notag\\
 +\left(d_1 \delta \phi_1 + d_{11} 3\phi_1^2\delta \phi_1 + d_{12}(\phi_2^2 \delta \phi_1 + 2 \phi_1 \phi_2 \delta \phi_2) -\delta \mu_1\right)\nu_1 \\
 +\left(d_2 \delta \phi_2 + d_{22} 3\phi_2^2\delta \phi_2 + d_{12}(\phi_1^2 \delta \phi_2 + 2 \phi_1 \phi_2 \delta \phi_1) -\delta \mu_2\right)\nu_2 d\omega \notag\\
= \int_{\Omega}- \left(\nabla \mu_1 , \nabla \psi_1 \right)
+ R_1(\phi_1,\phi_2) \psi_1      
- \left(\nabla \mu_2 , \nabla \psi_2 \right)	
+ R_2(\phi_1,\phi_2)\psi_2 \\
+(d_1 \phi_1 + d_{11} \phi_1^3 + d_{12} \phi_2^2 \phi_1 - \mu_1)\nu_1
+(d_2 \phi_2 + d_{22} \phi_2^3 + d_{12} \phi_1^2 \phi_2 - \mu_2)\nu_2 d\omega. 
\end{multline}
Space discretization is achieved by defining a regular grid with $N$ elements by direction, and using finite elements on the latter grid.
In Figure \ref{steadyStates}, we compare the steady states obtained with two different methods.
On the left column, we have plotted the density solutions of (\ref{NLsteady}). 
The initial guess of the Newton method was set to the initial condition of the temporal model (\ref{CI}),
and parameters were set to $N=420$ and $\epsilon = 1.e^{-11}$.
On the right column, we have plotted the density solutions obtained from simulating the temporal model (\ref{dGS}, \ref{mGS}) until onvergence, for the same sets of model parameters.
The same four cases described in the first paragraph were considered.
We observe a very good agrement between the two methods, despite the different approaches, and especially the different meshing techniques.
Direct steady state computation presents the advantage of being incomparably faster; 
however, as for any Newton method approach, convergence depends on a correct choice for the initial guess.
A posteriori error evaluation can be done by injecting the numerical solution $(\phi^1_h,\phi^2_h)$ into the original equation, 
and evaluating the residual:
\begin{equation}
R_h(\phi^1_h,\phi^2_h) = (\Delta(\mu_1(\phi^1_h,\phi^2_h)) + R_1(\phi^1_h,\phi^2_h), \Delta(\mu_2(\phi^1_h,\phi^2_h)) + R_2(\phi^1_h,\phi^2_h)).
\label{residual}
\end{equation}
This residual is null if and only if $(\phi^1_h,\phi^2_h)$ is a solution.

\begin{remark}
For the sake of claritity, only the case with two species has been presented.
Generalization to cases with $M \geq 2$ is straightforward, considering $W = (\phi_1,...,\phi_M,\mu_1,...,\mu_M)$.
\end{remark}

\section{Linear stability analysis}

In this section, we study the local stability of the steady states of systems (\ref{dGS}, \ref{mGS}).
First, using Fourier analysis, and then, using solving numerically a spectral problem.
For more details about the techniques used here, we refer the reader to \cite{Kuznetsov1998}, \cite{Manneville2004} and \cite{Tuckerman2004}.

\subsection{Fourier Analysis}

In this paragraph we assume that $\Omega = \mathbb{R}^2$.
The first step is to find homogeneous steady states of the diffusionless system, solution of:
\[
\begin{cases}
R_1 = 0, \\
R_2 = 0.
\end{cases}
\]
Then, we study the growth of a perturbation of the linearized system around a steady states using Fourier analysis.
Let us consider a perturbation $(\tilde{\phi_1},\tilde{\phi_2})$ around a steady state $(\bar{\phi_1},\bar{\phi_2})$ of the diffusionless system, and let us denote:
\begin{align*}
\phi_1(r,t) = \bar{\phi_1} + \tilde{\phi_1}(r,t),\\
\phi_2(r,t) = \bar{\phi_2} + \tilde{\phi_2}(r,t).
\end{align*}
We first linearise the reaction term as follows:
\[
\begin{pmatrix}
R_1(\phi_1,\phi_2) \\
R_2(\phi_1,\phi_2)
\end{pmatrix} 
\approx
\begin{pmatrix}
R_1(\bar{\phi_1},\bar{\phi_2}) \\
R_2(\bar{\phi_1},\bar{\phi_2})
\end{pmatrix} 
+ 
\mathbb{A}
\begin{pmatrix}
\tilde{\phi_1} \\
\tilde{\phi_2}
\end{pmatrix} 
,
\]
with:
\[
\mathbb{A}
=
\begin{pmatrix}
a_{11} & a_{12}\\
a_{21} & a_{22}\\
\end{pmatrix}
=
\begin{pmatrix}
\frac{\partial R_1}{\partial \phi_1} & \frac{\partial R_1}{\partial \phi_2}\\
\frac{\partial R_2}{\partial \phi_1} & \frac{\partial R_2}{\partial \phi_2}\\
\end{pmatrix}.
\]
Let us consider a perturbation of a self-diffusion terms acting on density $\phi_i$ ($i=1,2$):
\[
\Delta [(\bar{\phi}_i + \tilde{\phi_i})^3] 
=
\Delta [\bar{\phi}_i^3 + 3\bar{\phi}_i^2\tilde{\phi_i} + 3\bar{\phi}_i\tilde{\phi_i}^2 + \tilde{\phi_i}^3] 
\approx
\Delta \bar{\phi}_i^3 + 3\bar{\phi}_i\Delta \tilde{\phi_i},
\]
and a perturbation of a cross-diffusion term:
\begin{align*}
\Delta [(\bar{\phi}_1+\tilde{\phi_1})^2(\bar{\phi}_2+\tilde{\phi_2})] 
&=
\Delta [\bar{\phi}_1^2\bar{\phi}_2 
+ \bar{\phi}_1^2\tilde{\phi_2} 
+ 2\bar{\phi}_1\bar{\phi}_2\tilde{\phi_1} 
+ 2\bar{\phi}_1\tilde{\phi_1}\tilde{\phi_2}
+ \tilde{\phi_1}^2\bar{\phi}_2 + \tilde{\phi_1}^2\tilde{\phi_2}]\\
&\approx
\Delta (\bar{\phi}_1^2\bar{\phi}_i2) + 2\bar{\phi}_1\bar{\phi}_2\Delta(\tilde{\phi_1}) + \bar{\phi}_1^2\Delta(\tilde{\phi_2}).
\end{align*}
Using these identities to linearize the system, we may write the dynamics of the perturbation around an homogeneous equilibrium point:
\begin{equation}
\frac{\partial}{\partial t}
\begin{pmatrix}
\tilde{\phi}_1\\
\tilde{\phi}_2
\end{pmatrix}
=
\mathbb{A} 
\begin{pmatrix}
\tilde{\phi}_1\\
\tilde{\phi}_2
\end{pmatrix}
+
\mathbb{D} 
\begin{pmatrix}
\Delta (\tilde{\phi}_1)\\
\Delta (\tilde{\phi}_2)
\end{pmatrix}
,
\label{perturbation}
\end{equation}
with:
\[
\mathbb{D}
=
\begin{pmatrix}
\delta_{11} & \delta_{12}\\
\delta_{21} & \delta_{22}\\
\end{pmatrix}
=
\begin{pmatrix}
d_1 + 3d_{11}\bar{\phi_1}^2 + 2d_{12}\bar{\phi_1}\bar{\phi_2} & d_{12}\bar{\phi_1}^2 \\
d_{12}\bar{\phi_2}^2 & d_2 + 3d_{22}\bar{\phi_2}^2 + 2d_{12}\bar{\phi_1}\bar{\phi_2} \\
\end{pmatrix}.
\]
Classically, in reaction diffusion systems containing linear diffusion only, 
a Turing instability is an instability that appears around a stable homogeneous steady state, 
for which $\mbox{Tr}(\mathbb{A}) < 0$ and $\mbox{Det}(\mathbb{A}) > 0$, 
for a sufficiently high value of diffusion of one species (when the inhibitor diffuses faster than the activator). 
Assuming that $\mbox{Tr}(\mathbb{A}) < 0$, $\mbox{Det}(\mathbb{A}) > 0$,
one can adapt the proof to generalize it to nonlinear diffusion case, 
assuming that the coefficients of $\mathbb{D}$ are non-negative.
The Fourier transform of the perturbation equation (\ref{perturbation}) reads:
\[
\frac{\partial}{\partial t}
\begin{pmatrix}
\widehat{\tilde{\phi}}_1\\
\widehat{\tilde{\phi}}_2
\end{pmatrix} 
= 
\mathbb{A}_{\xi}
\begin{pmatrix}
\widehat{\tilde{\phi}}_1\\
\widehat{\tilde{\phi}}_2
\end{pmatrix}, 
\]
with:
\[
\mathbb{A}_{\xi} 
= \mathbb{A} - \|\xi\|^2\mathbb{D}
=
\begin{pmatrix}
a_{11} - \delta_{11}\|\xi\|^2 & a_{12} - \delta_{12}\|\xi\|^2\\
a_{21} - \delta_{21}\|\xi\|^2 & a_{22} - \delta_{22}\|\xi\|^2\\
\end{pmatrix}
.
\]
The perturbation analysis is reduced to the spectral analysis of matrix $\mathbb{A}_{\xi}$.
For any mode $\xi$, the trace is given by:
\[
\mbox{Tr}(\mathbb{A}_{\xi})
=
\mbox{Tr}(\mathbb{A}) - \|\xi\|^2\mbox{Tr}(\mathbb{D}),
\]
and the determinant by:
\[
\mbox{Det}(\mathbb{A}_{\xi})
= 
\mbox{Det}(\mathbb{D}) \|\xi\|^4 
+ \|\xi\|^2 (a_{21}\delta_{12} + \delta_{21}a_{12} - a_{11}\delta_{22} - \delta_{11}a_{22})
+ \mbox{Det}(\mathbb{A}).
\]
The eigenvalues of $\mathbb{A}_{\xi}$, roots of polynomial:
\[
\sigma^2 - \sigma \mbox{Tr}(\mathbb{A}_{\xi}) + \mbox{Det}(\mathbb{A}_{\xi}),
\]
are given by:
\[
\sigma_{\pm} = \frac{1}{2}\left(\mbox{Tr}(\mathbb{A_{\xi}}) \pm \sqrt{\Delta_{\xi}}) \right),
\]
with:
\[
\Delta_{\xi} = \left(\mbox{Tr}(\mathbb{A_{\xi}})\right)^2 - 4 \mbox{Det}(\mathbb{A}_{\xi}).
\]
A mode $\xi$ is stable if and only if $\mbox{Tr}(\mathbb{A_{\xi}}) < 0$ and $\mbox{Det}(\mathbb{A_{\xi}}) > 0$.
Indeed, the first condition ensures that at least one eigenvalue is negative, and the second ensures that both eigenvalues have the same sign.
Using the definition of $\mathbb{A}$ and $\mathbb{D}$, and the assumptions on the model, we know that $\mbox{Tr}(\mathbb{A}_{\xi}) <0$.
Therefore, stability analysis is reduced to the verification of the sign of $\mbox{Det}(\mathbb{A}_{\xi})$.
Let us rewrite:
\[
\mbox{Det}(\mathbb{A}_{\xi})
=
\mbox{Det}(\mathbb{D}) \|\xi\|^4 
- C \|\xi\|^2
+ \mbox{Det}(\mathbb{A})
=
\mbox{Det}(\mathbb{D})(\|\xi\|^2 - \alpha)^2 + \beta,
\]
with:
\begin{align*}
C &= -a_{21}\delta_{12} - \delta_{21}a_{12} + a_{11}\delta_{22} + \delta_{11}a_{22},\\
\alpha &= \frac{C}{2 \mbox{Det}(\mathbb{D})},\\
\beta &= \mbox{Det}(\mathbb{A}) - \mbox{Det}(\mathbb{D})\alpha^2.
\end{align*}
The function $\|\xi\|^2 \to \mbox{Det}(\mathbb{A_{\xi}})$ has a minimum at $\|\xi\|^2 = \alpha$, and we know that for $\|\xi\|^2=0$, 
$\mbox{Det}(\mathbb{A_{\xi}}) = \mbox{Det}(\mathbb{A}) >0$.
Let us remark that if $\alpha < 0$, there is no real root, therefore, as $\mbox{Det}(\mathbb{A})>0$, then $\mbox{Det}(\mathbb{D})>0$, the mode is stable.
If $\alpha >0$, the mode $\xi$ is unstable if the determinant is negative, that is if:
\[
\beta < 0
\iff
\mbox{Det}(\mathbb{A}) < \mbox{Det}(\mathbb{D})\alpha^2
\iff
2 \sqrt{\mbox{Det}(\mathbb{A}) \mbox{Det}(\mathbb{D})} < C.
\label{dispersion}
\]
This last inequality is the dispersion condition.
At the threshold of instability, we have:
\begin{equation}
C - 2\sqrt{\mbox{Det}(\mathbb{A}) \mbox{Det}(\mathbb{D})} = 0,
\label{dispersion}
\end{equation}
and only one mode $\xi$ is unstable, solution of:
\[
\beta = 0 \iff \alpha = \sqrt{\frac{\mbox{Det}(\mathbb{A})}{\mbox{Det}(\mathbb{D})}}.
\]
Therefore, the wave number is:
\[
\xi = \sqrt{\alpha} = \left( \frac{\mbox{Det}(\mathbb{A})}{\mbox{Det}(\mathbb{D})} \right)^{\frac{1}{4}},
\]
and the wavelength is given by:
\begin{equation}
\lambda = \frac{2\pi}{\xi}.
\label{wl}
\end{equation}

\begin{figure}[ht]\centering
\begin{subfigure}{0.6\textwidth}
\includegraphics[width=\linewidth]{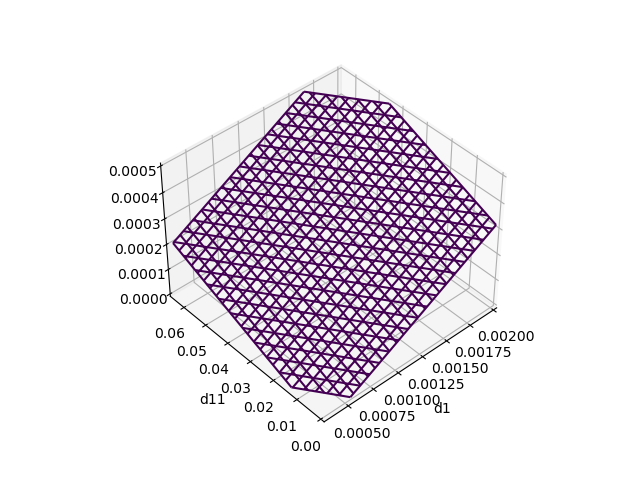}
\caption{Marginal stability surface, defined as the zero set of equation (\ref{dispersion}),
\\for $d_2 = 1e^{-5}$ and $d_{22}=0$.}
\end{subfigure}\\
\begin{subfigure}{0.4\textwidth}
\includegraphics[width=\linewidth]{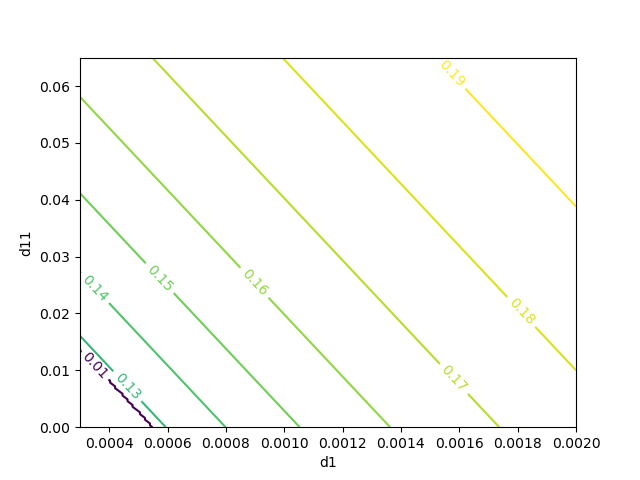}
\caption{Isolines of critical wavelength (formula (\ref{wl})): 
\\without cross-diffusion
\\($d_2 = 1e^{-5}$, $d_{22}=d_{12} = 0$.)}
\end{subfigure}
\begin{subfigure}{0.4\textwidth}
\includegraphics[width=\linewidth]{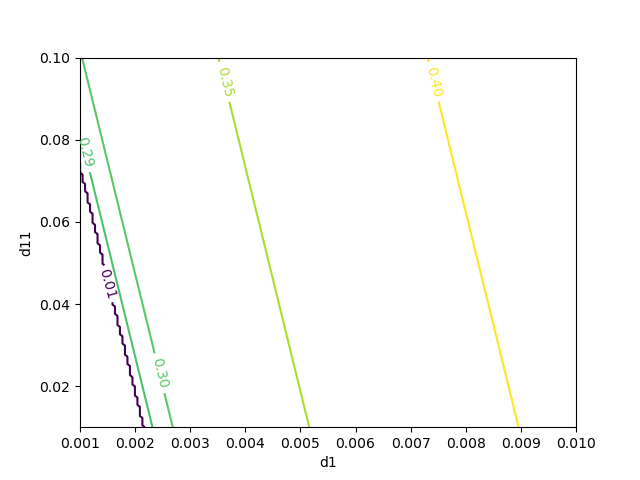}
\caption{Isolines of critical wavelength (formula (\ref{wl})): 
\\with cross-diffusion 
\\($d_2 = 1e^{-5}$, $d_{22} = 0$, $d_{12} = 4e^{-4}$).}
\end{subfigure}
\caption{Linear Stability Analysis: 
study of a linear perturbation around homogeneous steady state $(\phi_1^{-},\phi_2^{+})$ of \eqref{NLsteady}, 
in the case of the Gray-Scott reacting term with $F = 0.05$ and $k = 0.01$.
The marginal stability surface (on top), computed using formula \eqref{dispersion} indicates the stability threshold.
The isolines (on the bottom), computed using formula \eqref{wl}, give, for different plane cut of the stability surface ($d_{12}=0$, $d_{12} = 4e^{-4}$), 
the wavelength corresponding to the instability threshold.
For this set of parameters, we observe that cross-diffusion increases the critical wavelength significantly, 
and that it can not trigger instability without presence of linear and/or self diffusion.}
\label{figLSA}
\end{figure}

In this paragraph, we apply the previous results to the Gray-Scott system.
The reaction matrix reads:
\[
\mathbb{A}
=
\begin{pmatrix}
-F - \bar{\phi}_2^2  & -2\bar{\phi}_1\bar{\phi}_2 \\
\bar{\phi}_2^2  & 2\bar{\phi}_1\bar{\phi}_2 - (F+k)\\
\end{pmatrix}
.
\]
In particular, let us remark that $\mbox{Tr}(A) < 0$. Steady states are given by:
\[
\begin{cases}
\phi_1\phi_2^2 = F(1-\phi_1), \\
\phi_1\phi_2^2 = (F+k)\phi_2.
\end{cases}
\]
For any value of $F$ and $k$, a trivial solution exists:
\[
\begin{cases}
\bar{\phi}^0_1 = 1, \\
\bar{\phi}^0_2 = 0.
\end{cases}
\]
This steady state is always stable. Assuming that $(\phi_1,\phi_2) \not= (1,0)$, one may find:
\begin{align*}
\bar{\phi}_1^{\pm} &= \frac{1}{2} \left(1 \pm \sqrt{1 - 4\frac{(F+k)^2}{F}} \right),\\
\bar{\phi}_2^{\mp} &= \frac{F}{(F+k)}\bar{\phi}_1^{\pm}.
\end{align*}
However, these two equilibrium points $(\bar{\phi}_1^+,\bar{\phi}_2^-)$ and $(\bar{\phi}_1^-,\bar{\phi}_2^+)$ only exist if the square root 
$\sqrt{1 - 4\frac{(F+k)^2}{F}}$ is defined, that is when:
\[
k = -F + \frac{\sqrt{F}}{2} \mbox{ and } 0 \leq F \leq \frac{1}{4}.
\]
In other words, the diffusionless system undergoes a saddle-node bifurcation. 
Figure \ref{figLSA} shows the marginal stability surface, defined as the zero set of equation (\ref{dispersion}), 
around the unstable state $(\phi_1^-,\phi_2^+)$ with $k = 0.01$, $F = 0.05$, together with the critical wavelengths (formula (\ref{wl})).
Instability occurs at a threshold value, that depends linearly on different types of diffusion parameters ($d_1$, $d_{11}$, $d_{12}$).
We observe in particular that the presence of cross-diffusion increases significantly the critical wavelength, the other parameters being fixed.
We also observe, looking at the marginal stability surface around the origin, that the cross-diffusion term cannot trigger an instability alone, whereas linear diffusion can, but also self-diffusion.

\subsection{Spectral Analysis}

Classical LSA based on Fourier analysis may not be easily applied around non homogeneous steady states, as some convenient simplifications do not hold anymore.
However, a numerical analysis may be done, adapting the former method, leading to the evaluation of eigenvalues of a linear operator. 
Let us consider a perturbation $\delta U = (\delta \phi_1, \delta \phi_2)$ around a possibly non homogeneous steady state $\bar{U} = (\bar{\phi_1},\bar{\phi_2})$.
The dynamics of the perturbation may be written as:
\[
\frac{\partial}{\partial t} \delta U
=
\frac{\partial}{\partial t}(\bar{U} + \delta U)
= 
R(\bar{U} + \delta U)
\approx
\underbrace{R(\bar{U})}_{=0} + L(\bar{U})\delta U.
\]
Hence, linear stability analysis of steady state $\bar{U}$ is reduced to the analysis of the spectrum of linear operator $L(\bar{U})$.
One has to solve the eigenvalue problem:
\begin{equation}
L(\bar{U}) \delta U = \lambda \delta U.
\label{spectral}
\end{equation}
Using former linearization expression, one may write the epression of a linear operator $L$:
\[
L(\bar{U}) (\delta \phi_1, \delta \phi_2)
=
\begin{pmatrix}
\frac{\partial R_1}{\partial \phi_1}\delta \phi_1 
+ \frac{\partial R_1}{\partial \phi_2} \delta \phi_2 
+ d_1 \Delta \delta \phi_1 
+ 3d_{11}\Delta (\bar{\phi_1}^2\delta \phi_1) 
+ d_{12}(2\Delta(\bar{\phi_1}\bar{\phi_2}\delta \phi_2) + \Delta(\bar{\phi_2}^2\delta \phi_1))\\
\frac{\partial R_2}{\partial \phi_1}\delta \phi_1 
+ \frac{\partial R_2}{\partial \phi_2} \delta \phi_2 + 
d_2 \Delta \delta \phi_2 
+ 3d_{22}\Delta (\bar{\phi_2}^2\delta \phi_2) 
+ d_{12}(2\Delta(\bar{\phi_1}\bar{\phi_2}\delta \phi_1) + \Delta(\bar{\phi_1}^2\delta \phi_2))
\end{pmatrix}
\]
The variational formulation of the spectral problem is then to find eigenvalues $\lambda$ and eigenvectors $(\delta \phi^{\lambda}_1, \delta \phi^{\lambda}_2)$ such that:
\begin{align*}
\forall \theta_1, \theta_2 \in H^1(\Omega),
\int_{\Omega}
&- \left(d_1 \nabla(\delta \phi^{\lambda}_1)
+ 3 d_{11} \nabla(\bar{\phi_1}^2 \delta \phi^{\lambda}_1)
+ d_{12} (  2 \nabla(\bar{\phi_1}\bar{\phi_2}\delta \phi^{\lambda}_2) 
+ \nabla(\bar{\phi_2}^2\delta \phi^{\lambda}_1)\right).\nabla(\theta_1) )\\
&+ \left(\frac{\partial R_1}{\partial \phi_1}\delta \phi^{\lambda}_1 + \frac{\partial R_1}{\partial \phi_2} \delta \phi^{\lambda}_2\right) \theta_1\\
&- \left(d_2 \nabla(\delta \phi^{\lambda}_2)
+ 3 d_{22} \nabla(\bar{\phi_2}^2 \delta \phi^{\lambda}_2)
+ d_{12} (  2 \nabla(\bar{\phi_1}\bar{\phi_2}\delta \phi^{\lambda}_1) 
+ \nabla(\bar{\phi_1}^2\delta \phi^{\lambda}_2)\right).\nabla(\theta_2) )\\
&+ \left(\frac{\partial R_2}{\partial \phi_1}\delta \phi^{\lambda}_1 + \frac{\partial R_2}{\partial \phi_2} \delta \phi^{\lambda}_2 \right) \theta_2
d\omega
=
\lambda \int_{\Omega} \delta\phi^{\lambda}_1 \theta_1 + \delta\phi^{\lambda}_2 \theta_2 d\omega
\end{align*}
This problem was numerically solved in Freefem++ using Arnoldi method implemented via arpack++ (\cite{Lehoucq1998}), using P2 finite elements, with $50$ nodes per direction. 
Other discretizations have been tested (P1 elements with $420$ nodes per direction, P2 elements with $100$ nodes per direction), leading to the same numerical results.
The first 1000 eigenvalues have been computed, for each case.

\section{Pattern formation}

This section is a numerical exploration of pattern formation observed as steady states of system (\ref{dGS}),(\ref{mGS}).
A multi-parameter bifurcation is reported.
Two phenomenons have been observed, influencing the equilibrium dissipative structures: 
a dependancy to initial condition, and a dependancy to the geometry of the domain.

\subsection{Multi-parameter bifurcations}

\begin{figure}[ht]\centering
\begin{subfigure}{0.4\textwidth}
\includegraphics[width=\linewidth]{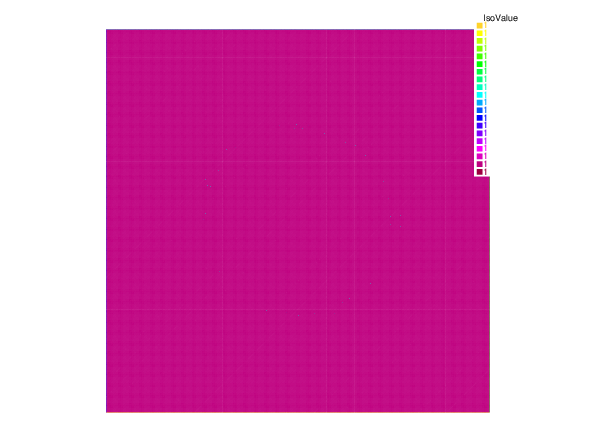}
\caption{Reference: stable case
\\$d_1 = 1e^{-5}, d_2 = 1e^{-5}$,
\\$d_{11} = d_{22} = d_{12}=0$.}
\end{subfigure}
\begin{subfigure}{0.4\textwidth}
\includegraphics[width=\linewidth]{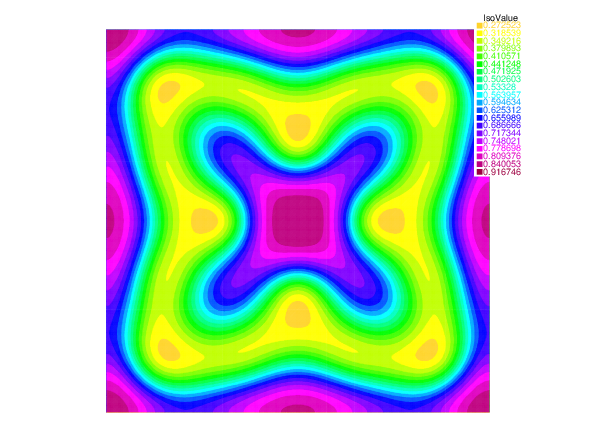}\\
\caption{Linear diffusion instability:
\\$d_1 = 2e^{-5}, d_2 = 1e^{-5}$,
\\$d_{11} = d_{22} = d_{12}=0$.}
\end{subfigure}
\begin{subfigure}{0.4\textwidth}
\includegraphics[width=\linewidth]{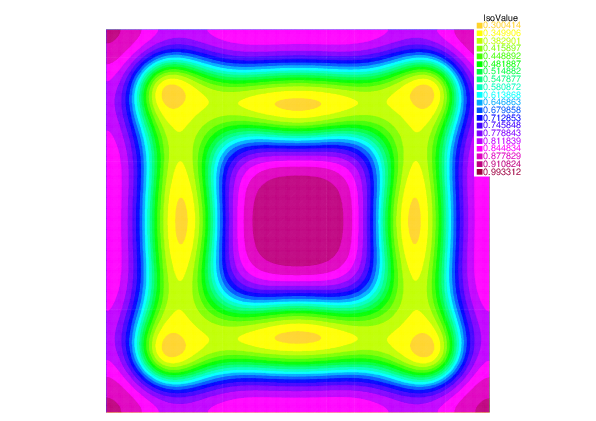}\\
\caption{Inhibitor self-diffusion instability:
\\$d_1 = 1e^{-5}, d_2 = 1e^{-5}$,
\\$d_{11} = 5e^{-6},d_{22} = d_{12} = 0$.}
\end{subfigure}
\begin{subfigure}{0.4\textwidth}
\includegraphics[width=\linewidth]{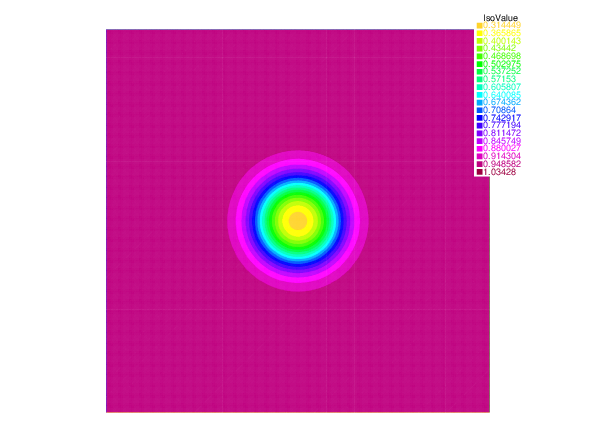}\\
\caption{Activator self-diffusion $d22$ instability:
\\$d_1 = 1.4e^{-5}, d_2 = 1e^{-5}$,
\\$d_{11} = 0, d_{22} = 2e^{-5}, d_{12}=0$.}
\end{subfigure}
\begin{subfigure}{0.4\textwidth}
\includegraphics[width=\linewidth]{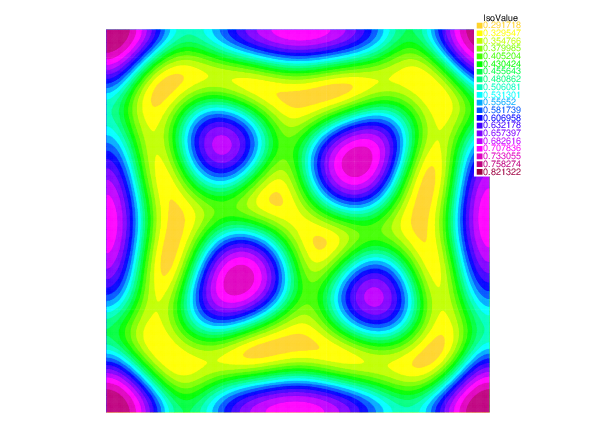}\\
\caption{Linear and self-diffusion instability:
\\$d_1 = 2e^{-5}, d_2 = 1e^{-5}$,
\\$d_{11} = 5e^{-6},d_{22} = d_{12}=0$.}
\end{subfigure}
\begin{subfigure}{0.4\textwidth}
\includegraphics[width=\linewidth]{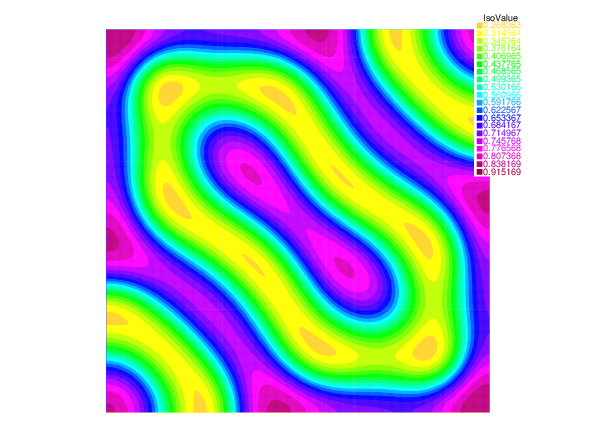}\\
\caption{Linear and cross-diffusion instability:
\\$d_1 = 2e^{-5}, d_2 = 1e^{-5}$,
\\$d_{11} = d_{22} = 0,d_{12} = 1e^{-6}$.}
\end{subfigure}
\caption{Numerical bifurcations: plot of density $\phi_1$ obtained by solving \eqref{NLsteady} for different sets of parameters, for the Gray-Scott reacting term, with $F = 0.037$ and $k = 0.06$.
We observe that four parameters can trigger a bifurcation towards different stable steady states.}
\label{figPT}
\end{figure}

On Figure \ref{figPT}, we study the case where $F=0.037$ and $k=0.06$. 
The Newton solver is initialized with \eqref{CI} and $C=0.3$, suppressing the random noise ($U=V=0$).
The domain $\Omega = [-0.2,0.2]^2$ is discretized with $N=420$ nodes per direction, 
defining a P1 element mesh on which the solution is computed.
In this case, only one homogeneous steady state exists, the trivial state $(1,0)$,
and Linear Stability Analysis predicts that this point is always stable.
Different combinations triggering an instability, leading to different dissipative structures, have been identified.
As expected, applying a sufficiently high linear diffusion leads to a classical Turing pattern (panel (b)).
More originally, the application of self-diffusion leads to the formation of dramatically different dissipative structures, 
depending on wether the self-diffusion is applied on the inhibitor $\phi_1$ (panel (c)), or the activator $\phi_2$ (panel (d)).
On panel (e), a linear diffusion is applied, sufficient to trigger an instability, and perturbed by cross-diffusion.
In this case, a diagonally oriented structure is observed, as opposed to the regular structure observed without cross-diffusion.
As seen on Figure \ref{figLSA}, cross-diffusion increases the critical wavelength:
the transition between vertical and diagonal orientation might be explained by the increase of the critical wavelength,
forcing the modes to grow diagonaly.
Let us remark that systems of the form (\ref{dGS}, \ref{mGS}) remain unchanged by applying a reflection of one of the axis, 
$R_x(x,y) = (-x,y)$ or $R_y(x,y) = (x,-y)$, or any rotation from the cyclic group $C_4$.
We conclude that the same density, after a mirror reflection, is also solution of the system, denoting the presence of at least three solutions in this case 
(one homogeneous, and two diagonally oriented). 
This could be the sign of an underlying pitchfork biurcation.
It would be interesting to carry further theoretical work on the problem.
On subfigure (f), a combination of linear and self- and cross-diffusion is applied.
As for the case of linear and cross-diffusion, a diagonally structure is observed; therefore, the same remarks apply.
For each case, the first $1000$ eigenvalues have been computed using \eqref{spectral}. 
They all have a negative real part, indicating linear stability.
We remark that the value of the first eigenvalue coresponds to the value of $-F$.

\begin{remark}
On Figure \ref{figPT}, 
the $L^1$ norm of the residual \eqref{residual} is typically $1e^{-3}$ for $\phi_1$, $1e^{-4}$ for $\phi_2$, 
the $L^{\infty}$ norm of the residual is of the order of $1e^{-2}$ for $\phi_1$, $1e^{-3}$ for $\phi_2$.
\end{remark}

\begin{center}
\begin{tabular}{|c|c|}
\hline
Case & Spectrum of $L$ (1000 first eigenvalues) \\
\hline
1 & $\lambda \in [-0.037,-0.515924]$ \\
2 & $\lambda \in [-0.037,-0.589005]$ \\
3 & $\lambda \in [-0.037,-0.537331]$ \\
4 & $\lambda \in [-0.037,-0.513677]$ \\
5 & $\lambda \in [-0.037,-0.618339]$ \\
6 & $\lambda \in [-0.037,-0.598155]$ \\
\hline
\end{tabular}
\end{center}

\subsection{Density effects}

\begin{figure}[ht]\centering
\begin{subfigure}{0.4\textwidth}
\includegraphics[width=\linewidth]{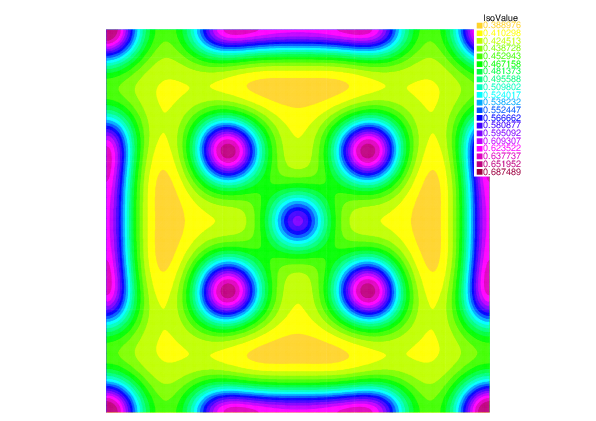}
\caption{Case 1: $C=0.3$ and $d_{22}=9e^{-5}$ \\and $\phi_2^0 = 0.25 \mathbf{1}_S$.}
\end{subfigure}
\begin{subfigure}{0.4\textwidth}
\includegraphics[width=\linewidth]{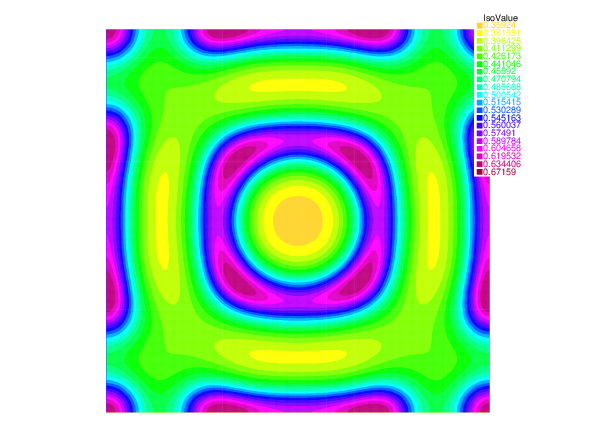}\\
\caption{Case 2: $C=0.5$ and $d_{22}=9e^{-5}$ \\and $\phi_2^0 = 0.25 \mathbf{1}_S$.}
\end{subfigure}
\begin{subfigure}{0.4\textwidth}
\includegraphics[width=\linewidth]{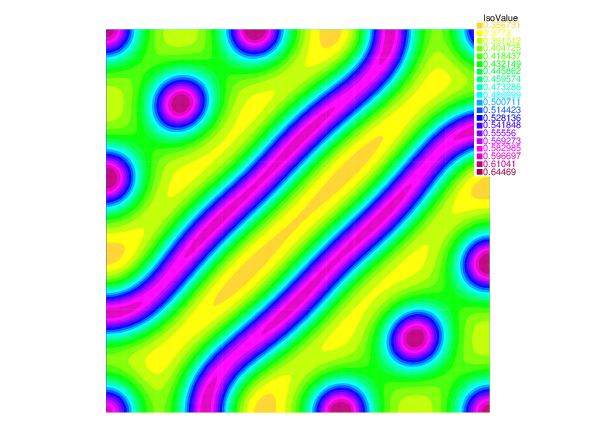}\\
\caption{Case 3: $C=0.3$ and $d_{22}=7.2e^{-5}$ \\and $\phi_2^0 = 0.25 \mathbf{1}_S$.}
\end{subfigure}
\begin{subfigure}{0.4\textwidth}
\includegraphics[width=\linewidth]{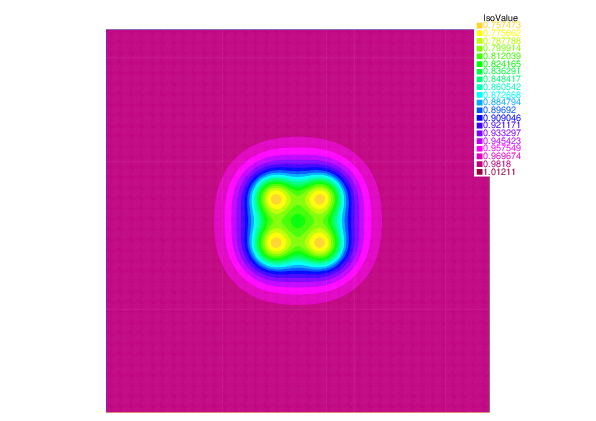}\\
\caption{Case 4: $C=0.3$ and $d_{22}=9e^{-5}$ \\and $\phi_2^0 = 0.15 \mathbf{1}_S$.}
\end{subfigure}
\caption{Plots of density $\phi_1$ obtained by computing \eqref{NLsteady} with Gray-Scott reacting term, 
with $F = 0.037$ and $k = 0.06$.
Parameters are set to ($d_1 = 2e^{-5}, d_2 = 0, d_{11} = 0, d_{12}=1e^{-6}$).
Initial condition and self diffusion are varied.
We observe a dependancy of the pattern with respect to the inital value of the density in the domain,
inducing a change of topology in the patterns.}
\label{figNC}
\end{figure}

In Figure \ref{figNC}, steady states for the system  (\ref{dGS}, \ref{mGS}) in the case of Gray Scott reaction is simulated, 
starting from different value of $C$ for initial conditions (\ref{CI}).
Three values of $C$ are tested: $0.3$, $0.41$ and $0.5$.
We observe that the final structure is dependant on the initial guess of the Newton method,
leading to clearly distinct dissipative patterns, differing notably by their topologies.
Reader may also note that the reaction term is the same for all tests
(in classical Gray Scott systems, one needs to vary the reaction term to obtain different patterns).
For a fixed value of $C$, decreasing the value of $d_{22}$ also modifies the final structure.
More generally, in the majority of the tests done preparing this article, decreasing the value of $d_{22}$,
and therefore increasing the ratio between cross and self-diffusion, 
lead to the formation of diagonally aligned dissipative patterns, after a threshold value. 
For each case, the first $1000$ eigenvalues have been computed using \eqref{spectral}. 
They all have a negative real part, indicating linear stability.
Here also, we remark that the value of the first eigenvalue coresponds to the value of $-F$.

\begin{center}
\begin{tabular}{|c|c|}
\hline
Case & Spectrum of $L$ (1000 first eigenvalues) \\
\hline
1 & $\lambda \in [-0.037,-0.476729]$ \\
2 & $\lambda \in [-0.037,-0.465432]$ \\
3 & $\lambda \in [-0.037,-0.433828]$ \\
4 & $\lambda \in [-0.037,-0.171386]$ \\
\hline
\end{tabular}
\end{center}

\subsection{Geometrical effects}

\begin{figure}[ht]\centering
\begin{subfigure}{0.4\textwidth}
\includegraphics[width=\linewidth]{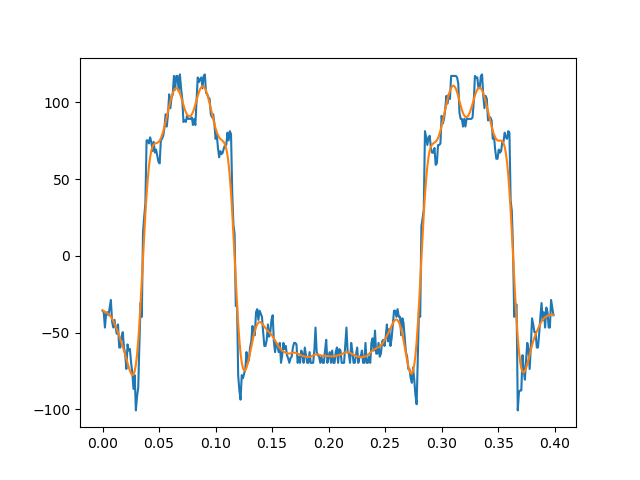}
\caption{Signal (horizontal cut at half-heigth), in blue, and low-pass filtered signal, in orange.}
\end{subfigure}
\begin{subfigure}{0.4\textwidth}
\includegraphics[width=\linewidth]{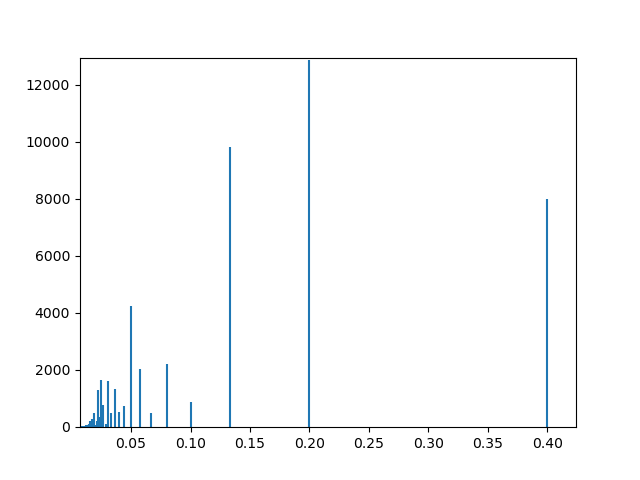}\\
\caption{Fourier transform of the signal}
\end{subfigure}
\caption{Fourier analysis of the steady state patterns of Figure \ref{figPT}, case (c).
We observe a peak around $0.2$, comparable in order of magnitude to the value obtained by linear stability analysis.}
\label{figFFT}
\end{figure}

In this paragraph, we study the effect of geometry on steady state solutions.
Figure \ref{figFFT} analyzes the pattern obtained in subfigure (c) of Figure \ref{figPT}.
We consider a horizontal cut at half height of the image, giving a 1d signal, represented in blue on subfigure (a),
together with its low pass filtered version, in orange, emphasizing the harmonics that are present in the signal.
The Fourier Transform is represented on subfigure (b);
on abscissa, the wavelength $\lambda$ is represented (with an expected cut-off at $0.4$, length of the domain).
The spectrum presents its highest peak at $\lambda = 0.2$.
This value is the wavelength of the main harmonic present in the equilibrium steady state, as seen as a standing wave.
We also observe that this value corresponds, in terms of order of magnitude, to the theoretical value of the unstable wavelength obtained on subfigure (b) of Figure \ref{figLSA},
even though Figure \ref{figLSA} was done for a different (but close) set of parameters.

\begin{figure}[ht]\centering
\begin{subfigure}{0.45\textwidth}
\includegraphics[width=\linewidth]{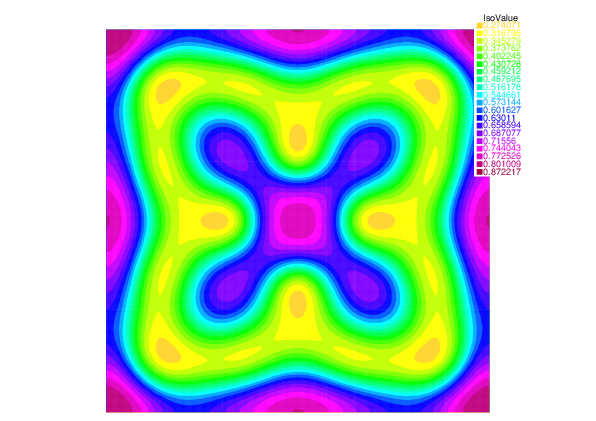}
\caption{Square: $L_x=L_y=0.4$.}
\end{subfigure}
\begin{subfigure}{0.45\textwidth}
\includegraphics[width=\linewidth]{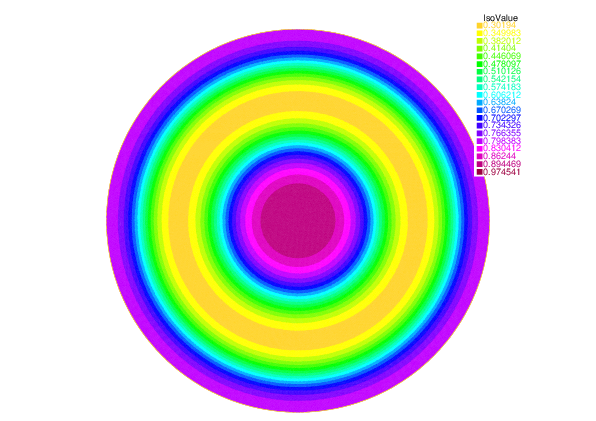}\\
\caption{Circle: radius $R=0.2$.}
\end{subfigure}
\begin{subfigure}{0.45\textwidth}
\includegraphics[width=\linewidth]{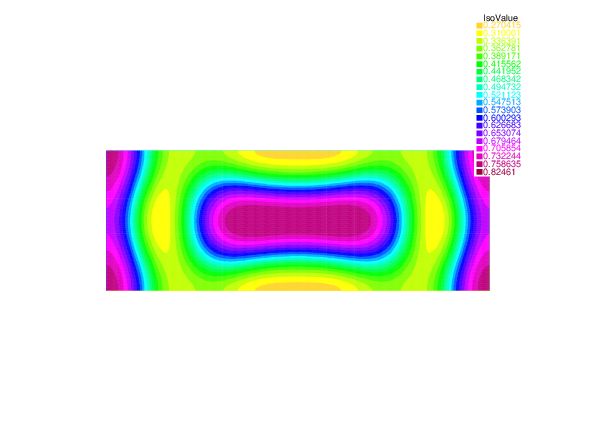}\\
\caption{Rectangle: $L_x = 0.41,L_y=0.15$.}
\end{subfigure}
\begin{subfigure}{0.45\textwidth}
\includegraphics[width=\linewidth]{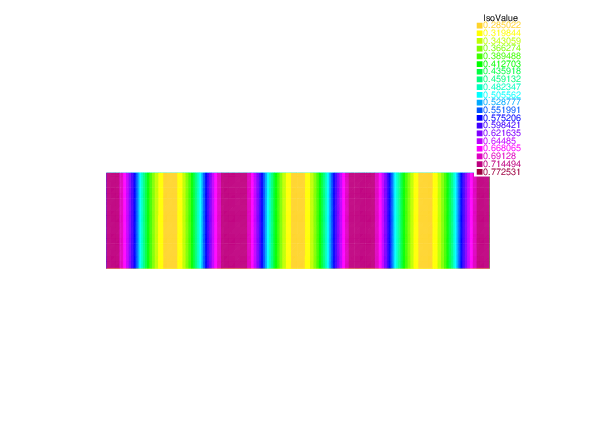}\\
\caption{Rectangle: $L_x = 0.4,L_y=0.1$.}
\end{subfigure}
\begin{subfigure}{0.45\textwidth}
\includegraphics[width=\linewidth]{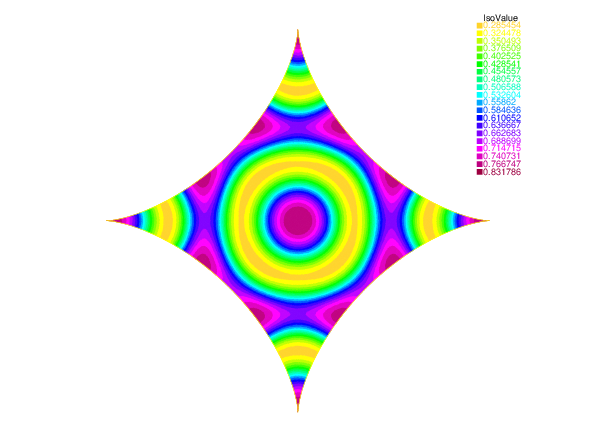}
\caption{Astroid: radius $R_{\max} = 0.3$.}
\end{subfigure}
\begin{subfigure}{0.45\textwidth}
\includegraphics[width=\linewidth]{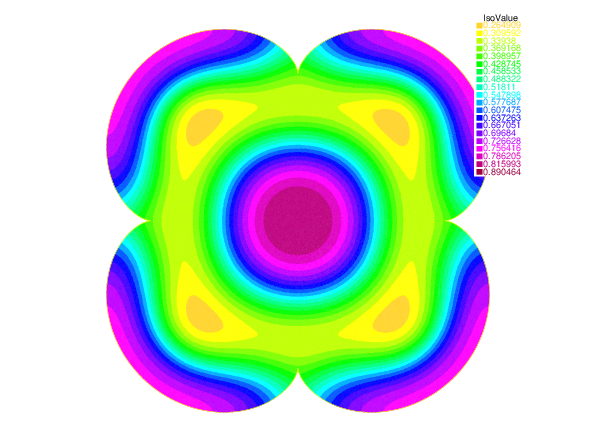}\\
\caption{Epicycloid: $R_{\max} = 0.12$.}
\end{subfigure}
\caption{Geometrical effects on steady state solutions
$\phi_1$ obtained by computing \eqref{NLsteady} with Gray-Scott reacting term, 
with $F = 0.037$ and $k = 0.06$, 
for a fixed set of parameters: $d_1 = 2e^{-5}, d_2 = 1e^{-5}, d_{11} = 2e^{-6}, d_{22} = 0, d_{12}=1e^{-6}$.
We observe a dependency of the patterns on the geometry of the domain.}
\label{figGeom}
\end{figure}

Figure \ref{figGeom} displays an analysis of the dependancy to steady state patterns with respect to the geometry of the domain in 2D.
The same set of parameters is used for all the tests, that is
$d_1=2e^{-5},d_{11}=2e^{-6},d_{12}=1e^{-6}, d_2=1e^{-5},d_{22}=0$, $C = 0.3$,
only the geometry is varied. For the geometries, we consider two rectangles ($Lx = 0.4, Ly = 0.1$, and $Lx = 0.41, Ly = 0.15$),
a circle of radius $R = 0.18$, 
an astroid: $(x(t),y(t)) = (R(3 \cos(t) + \cos(3t)), R(3. \sin(t) - \sin(3t)))$ with $R = 0.075$, 
and finally an epicycloid: $(x(t),y(t)) = (5R\cos(t)-R\cos(5t),5R\sin(t)-R\sin(5t))$ with $r = 0.03$.
In the cases of the circle, the astroid and the epicycloid, an initial mesh adaptation is done in order to ensure a almost constant space discretization,
with the value used for the square used as a reference $\Delta x = \frac{L}{N} = 0.00095$.
Let us consider the square geometry as the reference case (panel (a)).
Let us notice numerous symmetries, and the presence of spots, and a circling oscillating stripe. 
Moving to a circular case (panel (b)), all the spot patterns disappear, leading to a single, regular ring.
Comparing it with the 1D spectrum of Figure \ref{figFFT}, 
this means that only the main peak is present in the steady state, the other being cancelled by the geometry.
Decreasing the size of the square, one gets a transition phase (panel (c)), with a compressed version of the square pattern, until a threshhol value,
where a fully striped pattern is observed (panel (d)), favouring one direction only, where the instability can developp.
In the astroid case (panel (e)), two regimes may be identified:
a sufficient space allows the formation of a ring pattern, such as panel (b), as if there was only a circular border,
and sharp edges trigger the formation of stripes. 
This effect of curvature is similar in a sense to the nucleation process, favoured around sharp edges.
Inverting the curvature around the border singularities leads to the epicycloid case (panel (f)):
the central circular pattern is still present, and the inward cusps trigger a spot pattern in the latter.
This numerical exploration emphasizes the link between the critical wavelength and the geometry of the domain.
Further theoretical work could be done to investigate it,
as a close link between critical values of the dynamical system, 
and the eigenvalues and eigenvectors of the Laplace operator in $\Omega$ is suspected.
The first $1000$ eigenvalues of \eqref{spectral} have been computed, indicating linear stability.

\begin{center}
\begin{tabular}{|c|c|}
\hline
Case & Spectrum of $L$ (1000 first eigenvalues) \\
\hline
Circle              & $\lambda \in [-0.037,-0.763354]$ \\
Square              & $\lambda \in [-0.037,-0.608894]$ \\
Rectangle (large)   & $\lambda \in [-0.037,-1.50033]$ \\
Rectangle (thin)    & $\lambda \in [-0.037,-2.44972]$ \\
Epicycloid          & $\lambda \in [-0.037,-1.09816]$ \\
\hline
\end{tabular}
\end{center}

\section{Conclusion}

In this article, we have studied the long-term behavior of systems of the form (\ref{dGS}, \ref{mGS}).
First, a numerical investigation has been carried out, monitoring the energy evolution (\ref{Elaw}) in the temporal model (Figure \ref{Ecurve}),
with a Gray Scott reaction term. Energy dissipation and convergence to a steady state is almost always observed, 
except in the case where the system conveges to the homogeneous steady state.
An original method \eqref{Newton} has been proposed to directly solve the steady state model, and validated against the temporal model (Figure \ref{steadyStates}).
This method, like Newton methods in general, is very efficient, provided that a good initial guess is supplied.
However, several cases of divergence have been observed (an increase in mesh refinement may help, but this approach is limited).
As a second step, stability of the steady states solutions has been studied.
A classical linear stability analysis has been carried out around homogeneous steady states using Fourier analysis, 
leading to marginal stability surface (\ref{dispersion}) and critical wavelength (\ref{wl}) identification (Figure \ref{figLSA}).
In complement, a numerical method based on spectral analysis has been introduced \eqref{spectral} in order to study the stability of non-homogeneous steady states.
Finally, a numerical exploration has been done on equilibrium pattern formation.
In Figure \ref{figPT} a case has shown a rich dynamics, involving multi-parameters bifurcation, multistability and symmetry breaking (possibly histeresis).
On top of revealing original patterns (Figure \ref{figNC}), a link between dissipative structure and the initial energy injected in the system has been observed.
Also, a link between the domain geometry (Figures \ref{figFFT} and \ref{figGeom}), and the equilibrium wavelength of the steady states, has been reported.

As future works, several paths may be explored.
Improving the Newton method presented in the first section, and in particular its robustness to initial guess, could help to reach a broader variety of steady states solutions.
A potential mesh adaptation strategy could be an alternative, complementary way, to capture a broader range of solutions.
A systematic study of steady state patterns may be carried out, to identify sub groups in the parameter space.
The link between steady state spectrum, and domain geometry, may be investigated with the prism of spectral geometry, both theoretically and numerically.

\end{document}